\documentclass[aps,prb,twocolumn,superscriptaddress]{revtex4}

\usepackage{graphicx}
\usepackage{amsmath}
\usepackage{hyperref}
\hypersetup{colorlinks=true, citecolor=blue, urlcolor=blue, linkcolor=blue}

\newcommand{\ket}[1]{\left| #1 \right\rangle}
\newcommand{\bra}[1]{\left\langle #1 \right|}

\begin{document}
\graphicspath{}

\title{Band nesting, massive Dirac Fermions and Valley Lande and Zeeman effects in transition metal dichalcogenides: a tight--binding model}

\author{Maciej Bieniek}
\affiliation{Department of Physics, University of Ottawa, Ottawa, Ontario, Canada}
\affiliation{Department of Theoretical Physics, Wroc\l aw University of Science and Technology, Wybrze\.ze Wyspia\'nskiego 27, 50-370 Wroc\l aw, Poland}

\author{Marek Korkusi\'nski}
\affiliation{Department of Physics, University of Ottawa, Ottawa, Ontario, Canada}
\affiliation{Security and Disruptive Technologies, Emerging Technologies Division, NRC, Ottawa}

\author{Ludmi\l a Szulakowska}
\affiliation{Department of Physics, University of Ottawa, Ottawa, Ontario, Canada}

\author{Pawe\l ~Potasz}
\affiliation{Department of Physics, University of Ottawa, Ottawa, Ontario, Canada}
\affiliation{Department of Theoretical Physics, Wroc\l aw University of Science and Technology, Wybrze\.ze Wyspia\'nskiego 27, 50-370 Wroc\l aw, Poland}

\author{Isil Ozfidan}
\affiliation{D-Wave Systems, Vancouver, Canada}

\author{Pawe\l \ Hawrylak}
\affiliation{Department of Physics, University of Ottawa, Ottawa, Ontario, Canada}

\date{\today}

\begin{abstract}
We present here the minimal tight--binding model for a single layer of transition metal dichalcogenides (TMDCs) MX$_{2}$ (M--metal, X--chalcogen) which illuminates the physics and captures band nesting, massive Dirac Fermions  and Valley Lande and Zeeman magnetic field effects.  TMDCs share the hexagonal lattice with graphene but their electronic bands require much more complex atomic orbitals. Using symmetry arguments, a minimal basis consisting of 3 metal d--orbitals and 3 chalcogen dimer p--orbitals is constructed. The tunneling matrix elements between nearest neighbor metal and chalcogen orbitals are explicitly derived at $K$, $-K$ and $\Gamma$  points of the Brillouin zone.  The nearest neighbor tunneling matrix elements connect specific metal and sulfur orbitals  yielding an effective $6\times 6$ Hamiltonian giving correct composition of metal and chalcogen orbitals but not the direct gap at $K$ points. The direct gap at $K$, correct masses and conduction band minima at $Q$ points responsible for band nesting are obtained by inclusion of next neighbor Mo--Mo tunneling. The parameters of the next nearest neighbor model are successfully fitted to MX$_2$ (M=Mo, X=S) density functional (DFT) ab--initio calculations of the highest valence and lowest conduction band dispersion along $K-\Gamma$  line in the Brillouin zone. The effective two--band massive Dirac Hamiltonian for MoS$_2$, Lande g--factors and valley Zeeman splitting are obtained.
\end{abstract}


\maketitle

\section{Introduction}
There is currently renewed interest in understanding the electronic and optical properties of transition metal dichalcogenides (TMDCs) with formula MX$_2$ (M - metal from group IV to VI, X=S, Se, Te) \cite{R1, R2, R3, R4, R5, R6, R7, R8, R9, R10, R11, R12, R13, R14, R15, R16, R17, R18, R19, R20, R21, R22, R23, R24, R25, R26, R27, R28, R29}. Recent experiments and ab--initio calculations show that while bulk TMDCs are indirect gap semiconductors, single layers are direct gap semiconductors with direct gaps at $K$ points of the Brillouin zone \cite{R1, R2, R3, R4, R5, R6, R7, R8, R9, R10, R11, R12, R13, R14, R15, R16, R17, R18, R19, R20, R21, R22, R23, R24, R25, R26, R27, R28, R29}.  The existance of the gaps at $K$ points of the Brillouin zone (BZ) could be anticipated from graphene, as the two materials share the hexagonal lattice. If in graphene we were to replace one sublattice with metal atoms and second with chalcogen dimers, we might expect band structure similar to graphene but with opening of a gap at $K$ points in the BZ. If this analogy was correct, the gap opening in a spectrum of Dirac Fermions would lead to massive Dirac Fermions and nontrivial topological properties associated with broken inversion symmetry and valley degeneracy. However, in graphene the bandstructure can be understood in terms of a tight binding model with electrons tunneling between nearest neighbor's $p_z$ orbitals. The results of ab--initio calculations \cite{R2, R3, R4, R6, R7, R17, R20, R21, R23} for MX$_{2}$ show that the conduction band (CB) minima and valence band (VB) maxima wavefunctions are composed primarily of metal d--orbitals, i.e., next--nearest neighbors. If only metal orbitals are retained the lattice structure changes from hexagonal to triangular and the physics changes.  Additional complication is the presence of secondary conduction band minima at $Q$ points, at intermediate wavevectors between $K$ and $\Gamma$ points. These minima lead to conduction and valence band nesting which significantly enhances interactions of TMDCs with light \cite{R6, R17}. A tight--binding model which illuminates these aspects and allows for inclusion of magnetic field, confinement and many--body interactions is desirable. 

There are already several tight-binding approaches to TMDCs by, e.g., \textcite{R30}, \textcite{R31}, \textcite{R32}, \textcite{R33}, \textcite{R34} and others \cite{R35, R36, R37, R38, R39, R40, R56} as well as $k\cdot p$ approaches by \textcite{R41}. Each contribution brings new physics and adds on to our understanding of TMDCs. In this work we build on previous theoretical works as well as our ab--initio results \cite{R6, R23} to develop the simplest  tight--binding model which illuminates the physics of TMDCs, especially the role of hexagonal lattice, tunneling from metal to dimer orbitals, band nesting, effective two band massive Dirac Fermion model, Lande g--factors and valley Zeeman splitting and Landau levels.

\section{The model}
\begin{figure}
\includegraphics[width=0.40\textwidth]{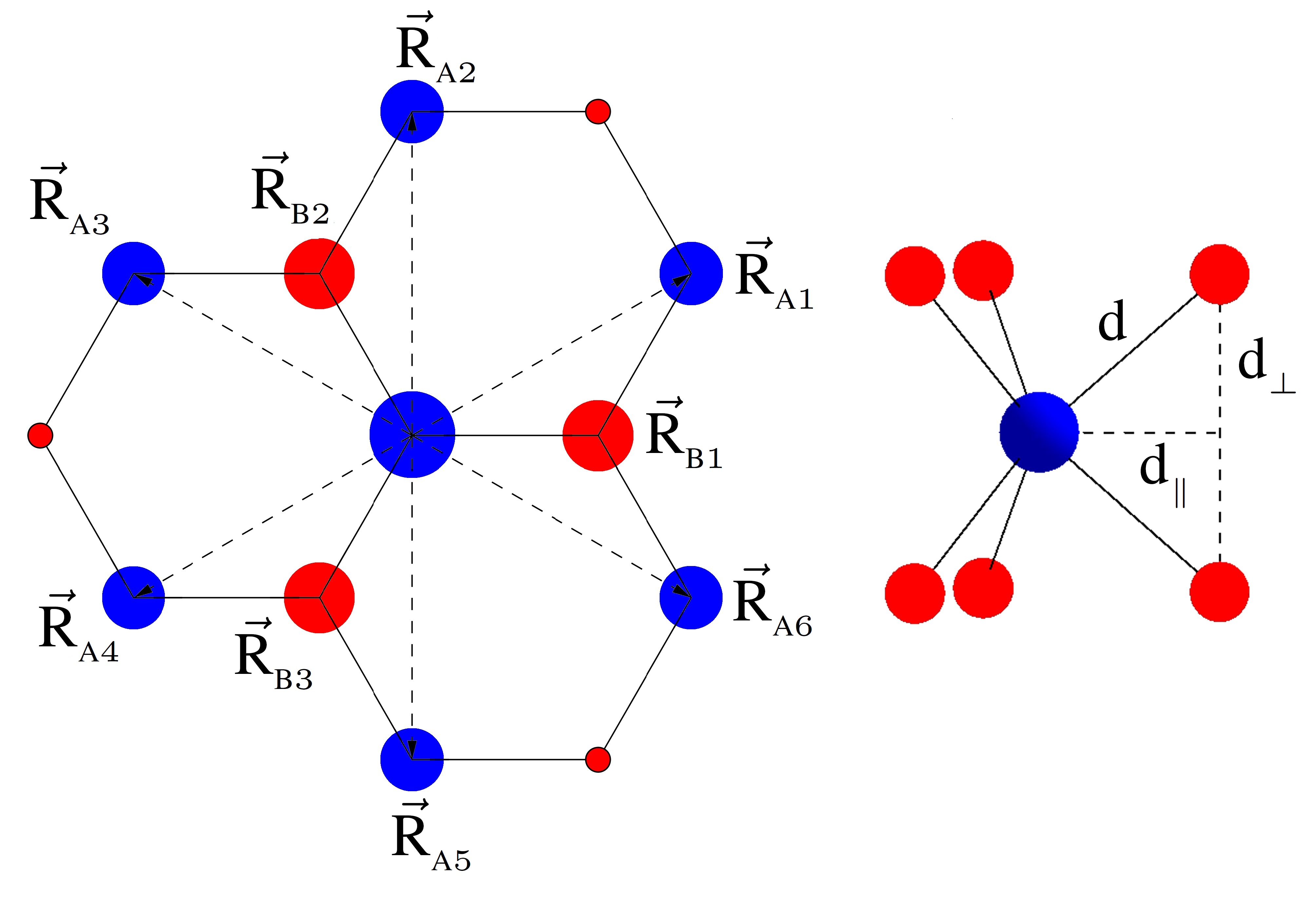}
\caption{\label{fig:F1} Structure of MX$_2$: blue Mo atoms, red sulfur atoms. Vectors $\vec{R}_{B1},\vec{R}_{B2},\vec{R}_{B3}$ point to nearest neighbors of central Mo atom; $\vec{R}_{A1}-\vec{R}_{A6}$ to next nearest neighbors of Mo atom.}
\end{figure}
We start with the structure of a single layer of MX$_2$ and for definiteness we focus on MoS$_2$. Fig.~\ref{fig:F1} shows the top view of a fragment of MoS$_2$ hexagonal lattice, with Mo positions marked with blue circles and sulfur dimers marked with red circles. The lattice structure is almost identical to graphene, the differences are visible in the side view showing the sulfur dimers and three, sulfur - metal - sulfur, layers of a single layer of MoS$_2$. Fig.~\ref{fig:F1} shows a central metal atom (large blue circle) of sublattice A surrounded by three first neighbor sulfur dimers of sublattice B marked with positions $\vec{R}_{B1}$, $\vec{R}_{B2}$ and $\vec{R}_{B3}$. The positions of second neighbors belonging to metal sublattice A are marked with $\vec{R}_{A1}$,...,$\vec{R}_{A6}$. We now construct the wavefunction out of orbitals localized on metal atoms and sulfur dimers. We start by selecting orbitals on a metal atom. Guided by results of ab--initio calculations \cite{R6} we first consider d--orbitals $l{=}2$,with $m_{_{d}} {=} {\pm}2,{\pm}1, 0$. Out of 5 $m_{_{d}}$  orbitals, orbitals with $m_{_{d}}{=}{\pm}2,0$ are even with respect to the Mo layer. We select three d--orbitals $\varphi_{l{=}2,m_{_{d}}}(\vec{r}-\vec{R}_{A,i})$ localized on i-th Mo atom of sublattice A at $\vec{R}_{A,i}$. For a sulfur dimer we select 3 p orbitals with $l{=}1, m_{_{p}}{=}{\pm} 1, 0$ on lower (L) and upper (U) sulfur atoms. We first construct dimer orbitals which are even with respect to the Mo plane: 
\begin{displaymath}
\varphi_{l=1,m_{_{p}}=\pm 1}\left(\vec{r}\right)=\frac{1}{\sqrt{2}}\left[\varphi_{l=1,m_{p}=\pm 1}^{U}\left(\vec{r}\right)+\varphi_{l=1,m_{p}=\pm 1}^{L}\left(\vec{r}\right)\right]
\end{displaymath}
and 
\begin{displaymath}
\varphi_{l=1,m_{_{p}}=0}\left(\vec{r}\right)=\frac{1}{\sqrt{2}}\left[\varphi_{l=1,m_{p}=0}^{U}\left(\vec{r}\right)-\varphi_{l=1,m_{p}=0}^{L}\left(\vec{r}\right)\right].
\end{displaymath}
We note the minus sign in the $m_{_{p}}{=}0$ orbital due to odd character of $m{=}0\ p_{z}$ orbital. With 3 orbitals on Mo atom we can write the wavefunctions on the sublattice A for each wavevector $\vec{k}$ and orbital $m_{_{d}}$ as:
\begin{equation}
\Psi_{A,m_{_{d}}}^{\vec{k}}\left(\vec{r}\right)=\frac{1}{\sqrt{N_{UC}}}\sum_{i=1}^{N_{UC}}e^{i\vec{k}\cdot\vec{R}_{A,i}}\varphi_{l=2,m_{_{d}}}\left(\vec{r}-\vec{R}_{A,i}\right)
\label{eq1}
\end{equation}
where $N_{UC}$ is number of unit cells. In the same way we can write the three wavefunctions for sublattice B of sulfur dimers:
\begin{equation}
\Psi_{B,m_{_{p}}}^{\vec{k}}\left(\vec{r}\right)=\frac{1}{\sqrt{N_{UC}}}\sum_{i=1}^{N_{UC}}e^{i\vec{k}\cdot\vec{R}_{B,i}}\varphi_{l=1,m_{_{p}}}\left(\vec{r}-\vec{R}_{B,i}\right)
\label{eq2}
\end{equation}
We now seek the LCAO electron wavefunction $\Psi_{n}^{\vec{k}}\left(\vec{r}\right)=\left[\sum_{m_{_{d}}}A_{m_{_{d}}}^{\vec{k}}(n)\Psi_{A,m_{_{d}}}^{\vec{k}}\left(\vec{r}\right)+\sum_{m_{_{p}}}B_{m_{_{p}}}^{\vec{k}}(n)\Psi_{B,m_{_{p}}}^{\vec{k}}\left(\vec{r}\right)\right]$  with coefficients $A_{m_{_{d}}}^{\vec{k}}(n)$, $B_{m_{_{p}}}^{\vec{k}}(n)$ for band $"n"$ and wavevector $\vec{k}$ to be obtained by diagonalizing the Hamiltonian matrix in the space of wavefunctions $\Psi_{A,m_{_{d}}}^{\vec{k}}\left(\vec{r}\right)$ and $\Psi_{B,m_{_{p}}}^{\vec{k}}\left(\vec{r}\right)$.

\section{The nearest-neighbor tunneling Hamiltonian}
We now proceed to construct matrix elements $\bra{\Psi_{A,m_{_{d}}}^{\vec{k}}}\hat{H}\ket{\Psi_{B,m_{_{p}}}^{\vec{k}}}$ of the Hamiltonian describing tunneling from Mo orbitals to sulfur dimer orbitals. The matrix elements for tunneling from Mo atom in Fig.~\ref{fig:F1} to it's 3 nearest--neighbors $\vec{R}_{B1}$, $\vec{R}_{B2}$ and $\vec{R}_{B3}$ can be explicitly written in analogy to graphene:
\begin{equation}
\begin{split}
&\bra{\Psi_{A,m_{_{d}}}^{\vec{k}}}\hat{H}\ket{\Psi_{B,m_{_{p}}}^{\vec{k}}}=\int d\vec{r} \varphi_{l=2,m_{_{d}}}^{*}\left(\vec{r}\right)V_{A}\left(\vec{r}\right)\cdot \\
&\bigg[ e^{i\vec{k}\cdot\vec{R}_{B1}}\varphi_{l=1,m_{_{p}}}\left(\vec{r}-\vec{R}_{B1}\right)+e^{i\vec{k}\cdot\vec{R}_{B2}}\varphi_{l=1,m_{_{p}}}\left(\vec{r}-\vec{R}_{B2}\right)+\\
&e^{i\vec{k}\cdot\vec{R}_{B3}}\varphi_{l=1,m_{_{p}}}\left(\vec{r}-\vec{R}_{B3}\right) \bigg],
\end{split}
\label{eq3}
\end{equation}
where $V_A(r)$ is a potential on sublattice A. We can evaluate matrix elements, Eq.~\ref{eq3}, at the $K$ point of the Brillouin zone $\left(K=\left[0,4\pi/\left(3\sqrt{3}d_{||}\right)\right]\right)$ to obtain
\begin{equation}
\begin{split}
&\bra{\Psi_{A,m_{_{d}}}^{\vec{k}=K}}\hat{H}\ket{\Psi_{B,m_{_{p}}}^{\vec{k}=K}}=\\
&\left(1+e^{i\left(1-m_{_{d}}+m_{_{p}}\right)2\pi/3}+e^{i\left(1-m_{_{d}}+m_{_{p}}\right)4\pi/3} \right)V_{pd}\left(m_{_{d}},m_{_{p}}\right)
\label{eq4}
\end{split}
\end{equation} 
where $V_{pd}(m_{_{d}},m_{_{p}})$ is a Slater--Koster matrix element for tunneling from Mo atom orbital $m_{_{d}}$ to nearest sulfur dimer orbital $m_{_{p}}$. We see in Eq.~\ref{eq4} that tunneling from central Mo atom to three nearest neighbor sulfur dimers generates additional phase factors which depend on the angular momentum of orbitals involved.  The pairs of orbitals giving non--vanishing tunneling matrix element must satisfy selection rule $1+m_{_{p}}-m_{_{d}}=0, \pm3$. The only pairs of orbitals which satisfy this rule at $K$ point are:
\begin{equation}
\left[m_{_{d}}{=}0, m_{_{p}}{=}{-}1 \right],\left[m_{_{d}}{=}2, m_{_{p}}{=}1 \right], \left[m_{_{d}}{=}{-}2, m_{_{p}}{=}0 \right].
\label{eq5}
\end{equation}
Hence the Hamiltonian at the $K$ point is block--diagonal. Similar calculations lead to different selection rules at the nonequivalent $-K$ point:
\begin{equation}
\left[m_{_{d}}{=}0, m_{_{p}}{=}1 \right],\left[m_{_{d}}{=}2, m_{_{p}}{=}{-}1 \right], \left[m_{_{d}}{=}{-}2, m_{_{p}}{=}1 \right],
\label{eq6}
\end{equation}
while at the  $\Gamma$ point different pairs of orbital are coupled: 
\begin{equation}
\left[m_{_{d}}{=}0, m_{_{p}}{=}0 \right],\left[m_{_{d}}{=}2, m_{_{p}}{=}{-}1 \right], \left[m_{_{d}}{=}{-}2, m_{_{p}}{=}1 \right].
\label{eq7}
\end{equation}
We see that the three $m_{_{d}}$ orbitals are coupled to a different $p$ dimer orbital each. Which pairs are coupled depends on the $K$ and $\Gamma$ points. This has important consequences for the response to magnetic field discussed later.  
We can now write tunneling Hamiltonian with first-- nearest--neighbor tunneling only. Here we put together the group of 3 degenerate d orbitals of Mo and a group of three degenerate p-orbitals of S$_2$. The tunneling matrix elements depend on tunneling amplitudes $V_{i}$ with dependence on $\vec{k}$ expressed by functions $f_{i}(\vec{k})$. The function $f_{0}(\vec{k})$ is the only function finite at $K=\left[0,4\pi/\left(3\sqrt{3}d_{||}\right)\right]$. Looking at the tunneling matrix elements of Hamiltonian, Eq.~\ref{eq8}, containing $f_{0}(\vec{k})$ gives the coupled  pairs of orbitals given by Eq.~\ref{eq5}. Explicit forms of $f_{i}(\vec{k})$ and $V_{i}$ are given in the Appendix A.
\begin{widetext}
\begin{eqnarray}
H\left(\vec{k} \right)=\left(\begin{array}{cccccc}
E_{m_{_{d}}=-2}& 0 & 0 & V_{1}f_{-1}(\vec{k}) & -V_{2}f_{0}(\vec{k}) & V_{3}f_{1}(\vec{k}) \\
& E_{m_{_{d}}=0} & 0   & -V_{4}f_{0}(\vec{k}) & -V_{5}f_{1}(\vec{k}) & -V_{4}f_{-1}(\vec{k}) \\
&  & E_{m_{_{d}}=2}    & -V_{3}f_{1}(\vec{k}) & -V_{2}f_{-1}(\vec{k}) & V_{1}f_{0}(\vec{k}) \\
&  &     & E_{m_{_{p}}=-1} & 0 & 0 \\
&  &     &  & E_{m_{_{p}}=0} & 0 \\
&  &     &  &  & E_{m_{_{p}}=1} \\
\end{array} \right)
\label{eq8}
\end{eqnarray}
\end{widetext}

\begin{figure}
\includegraphics[width=0.40\textwidth]{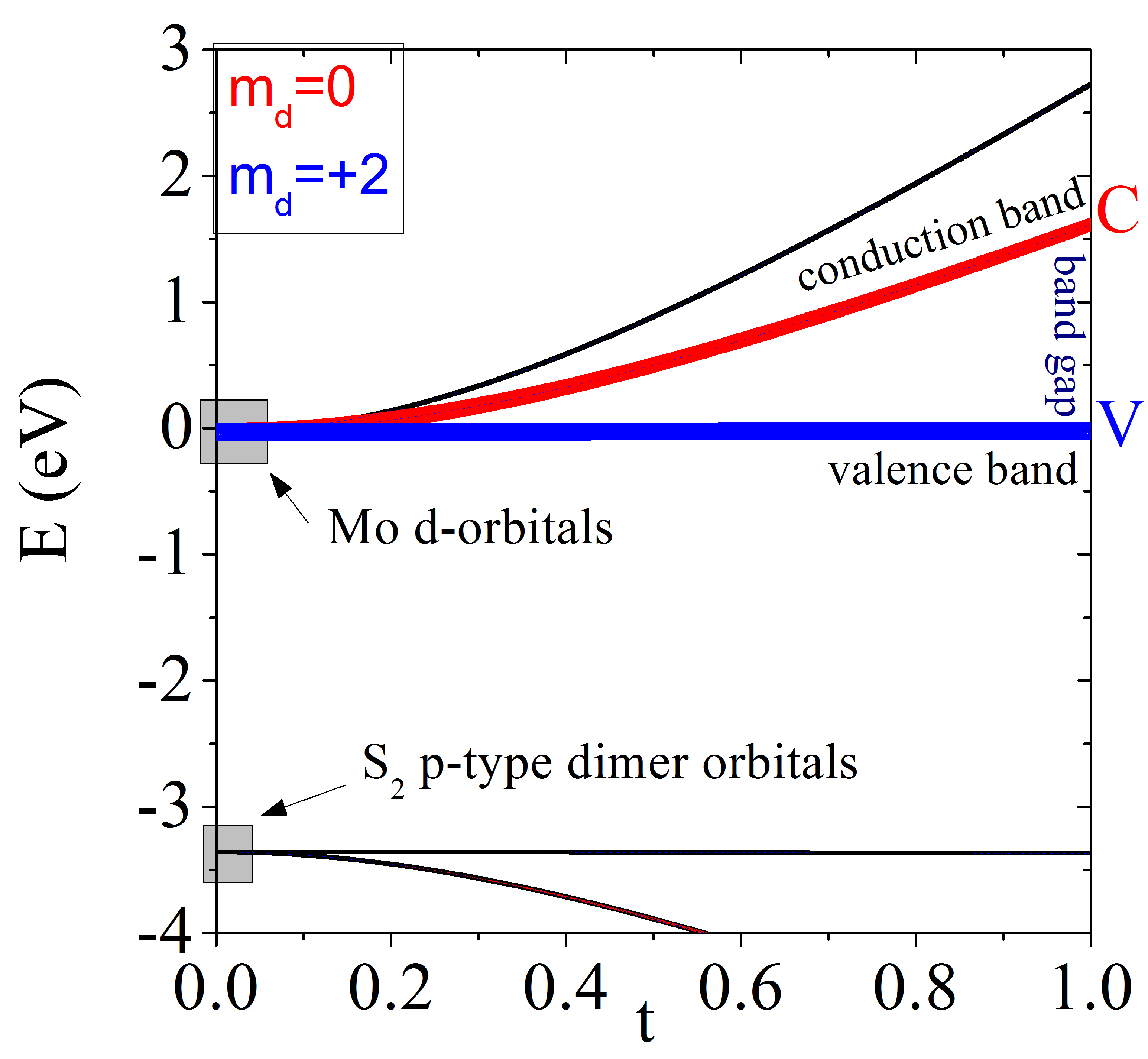}
\caption{\label{fig:F2} Evolution of eigenenergies at K point as a function of tunneling matrix element $t$ between Mo $d$ and S$_{2}$ dimer $p$ orbitals. Orbital composition -- red circles $m_{_{d}}{=}\ 0$, blue circles $m_{_{d}}{=}+2$. Conduction band C and valence band V are marked. Lower energy orbitals are S$_{2}$ $p$  orbitals. The removal of degeneracy of $d$ orbitals and opening of the gap between $m_{_{d}}{=}\ 0$ and $m_{_{d}}{=}+2$ orbitals is shown. All energies are measured from the top of the valence band V.}
\end{figure}
We parameterize tunneling matrix elements, $H_{ij}= t_{ij}$, of Eq.~\ref{eq8} with tunneling parameter $t$. $t{=}0$ means no tunneling and $t{=}1$ means full tunneling matrix, Eq.~\ref{eq8}. Fig.~\ref{fig:F2} shows the evolution of the energy spectrum of the first--nearest--neighbor Hamiltonian at K point, Eq.~\ref{eq8}, as a function of tunneling strength $t$. At $t{=}0$ we have 3 degenerate $d$-orbitals with energies $E_{_{d}}$ and 3 degenerate $p$-orbitals on sulfur dimers with energy $E_{_{p}}$. As the tunneling from Mo to S$_2$ orbitals is turned on the degeneracy of $d$-orbitals is removed as they start hybridizing with $p$-orbitals. The orbital $m_{_{d}}{=}2$ is the lowest energy valence band orbital. The $m_{_{d}}{=}0$ evolves as a conduction band orbital and $m_{_{d}}{=}-2$ gives rise to the higher energy conduction band orbital.  The magnitude of the bandgap is fitted to the ab--initio result using ABINIT and ADF \cite{R6,R23}.
\begin{figure}
\includegraphics[width=0.40\textwidth]{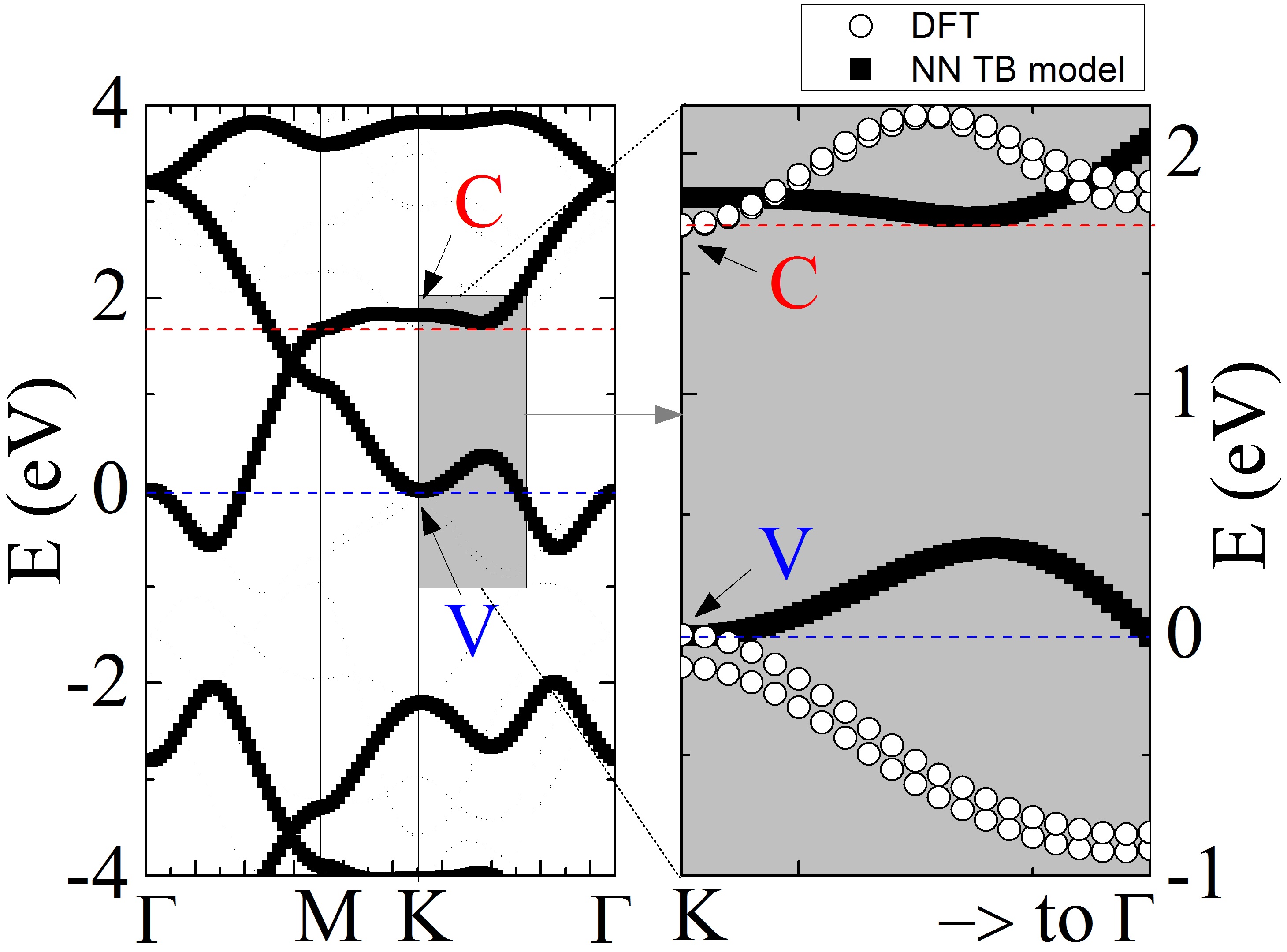}
\caption{\label{fig:F3} Nearest neighbor tunneling model. Black -TB model, white circles – DFT (no SO).  Left: Energy bands from  $K$ to $\Gamma$, $K$ to $M$ and $M$ to $\Gamma$ points in the Brillouinn zone obtained for MoS$_{2}$ with only nearest neighbor tunneling included. Note band gap closing along $M-\Gamma$ direction due to different symmetry of Mo orbitals at $K$ and $\Gamma$.  Right: Comparison of C and V bands close to $K$ point. Note local energy gap at K point but incorrect masses leading to lowest energy gap between $K$ and $\Gamma$ points.}
\end{figure}
Fig.~\ref{fig:F3} shows the energy bands across the BZ obtained by fitting the first neighbor Hamiltonian, Eq.~\ref{eq8}, using genetic algorithm to ab--initio results obtained using ABINIT \cite{R6,R23}. We see that such a simple Hamiltonian predicts a correct, finite, gap at $K$ point but it also predicts closing of the gap in the Brillouin zone, here shown between $M$ and $\Gamma$  points. The closing of the gap is a consequence of the reversal of the role of $m_{_{d}}{=}0$ d--orbital:  it  is a conduction band orbital at $K$ point but valence band orbital at $\Gamma$. Therefore without level repulsion there must be closing of the gap. In the right panel we also show close up of the dispersion of valence and conduction band along the $K-\Gamma$  line. We see that the gap at $K$ point is correct but the masses of holes and electrons are incorrect, leading to the lowest energy gap away from the $K$ point and a lack of CB maximum at the $Q$ point. Hence the simplest nearest-neighbor tunneling model which successfully describes Dirac Fermions in graphene captures the opening of the gap at $K$ point of the BZ and composition of VB and CB wavefunctions in terms of $d$--Mo and $p$--S$_{2}$ orbitals. However, it fails to capture important properties of CB and VB away from the $K$ points. In order to capture the effective masses of CB and VB bands and CB maximum leading to band nesting we need to include tunneling between second neighbor Mo atoms.
\section{The first and second neighbor tunneling Hamiltonian}
\begin{figure}
\includegraphics[width=0.40\textwidth]{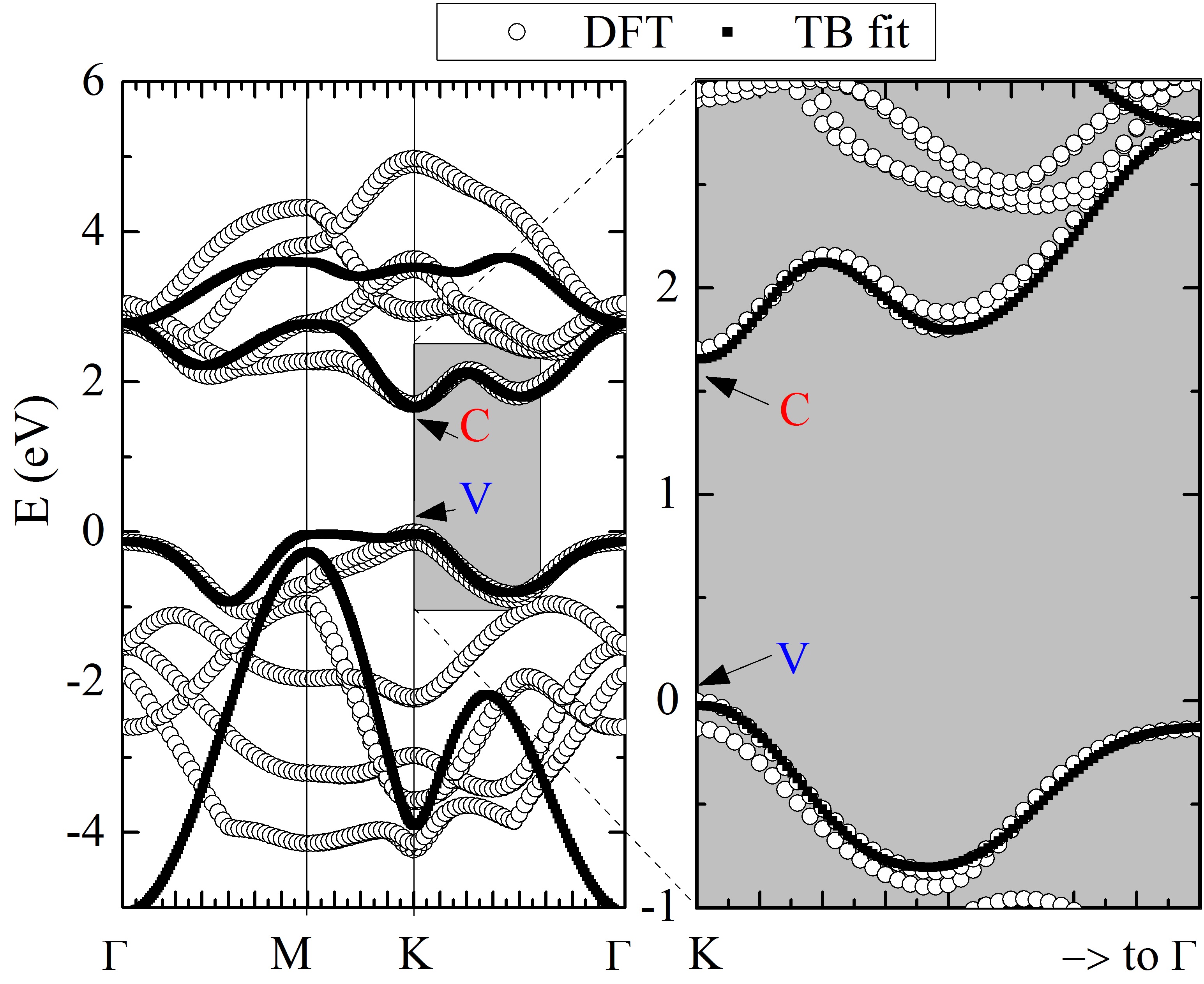}
\caption{\label{fig:F4} First and second neighbor tunneling model. Black --nn tb model, white circles -- DFT (no SO).  Left: Energy bands from $K$ to $\Gamma$, $K$ to $M$ and $M$ to $G$ points in the Brillouinn zone obtained for Mo--S$_2$ first and second neighbor tunneling model. Note gap opened across entire BZ. Right: Comparison of C and V bands close to $K$ point. Note direct energy gap at $K$ point, correct masses and CB minimum at $Q$ point.}
\end{figure}

\begin{figure}
\includegraphics[width=0.40\textwidth]{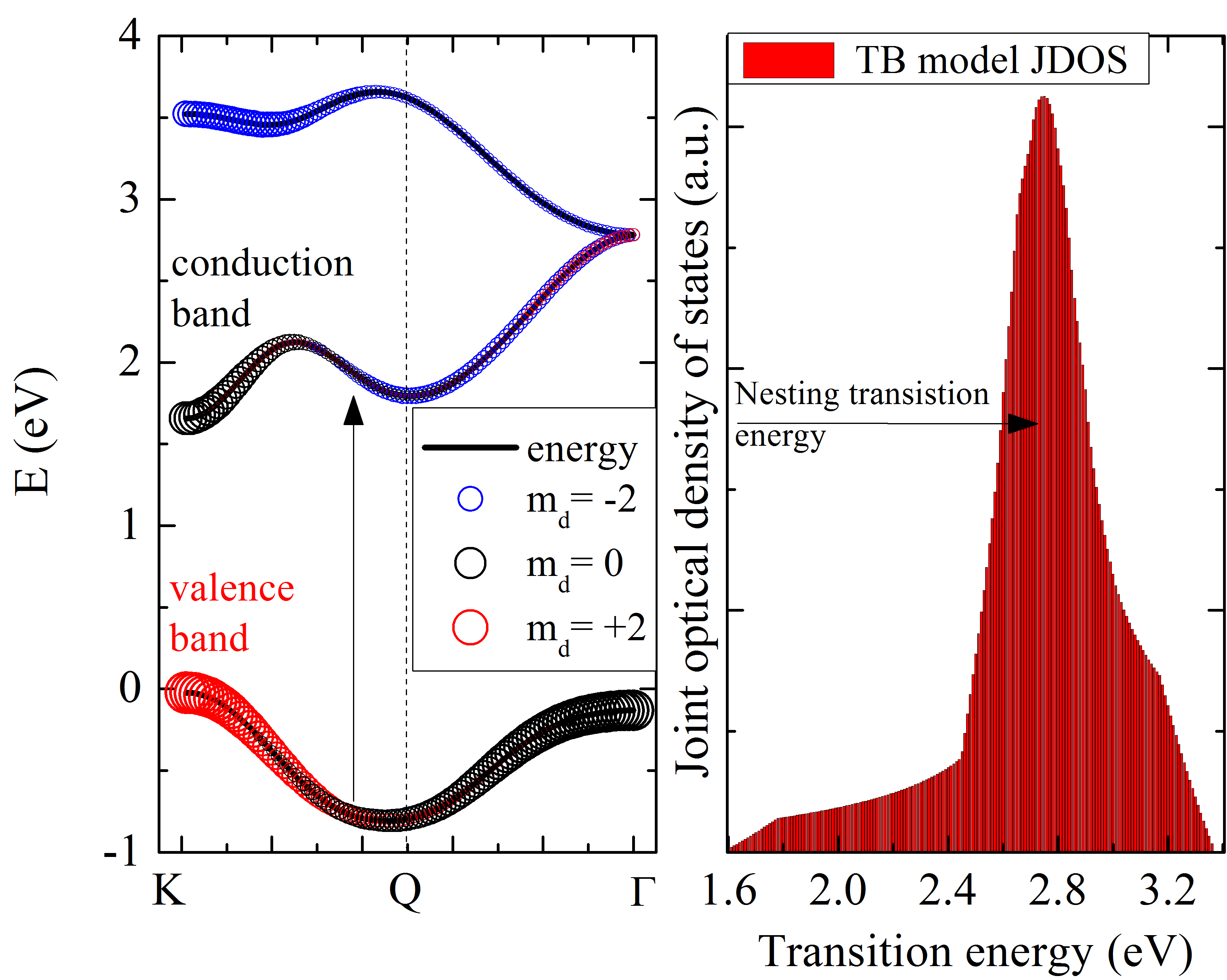}
\caption{\label{fig:F5} Origin of $Q$ point minima in CB and its effect on joint optical density of states. Left: Evolution of energy bands from $K$ to $Q$ and from $Q$ to $\Gamma$ points  in the Brillouin zone. In CB note $m_{_{d}}{=}0$ contribution at $K$ and $m_{_{d}}{=}{-}2$ at $Q$ while in VB note $m_{_{d}}{=}{+}2$ at $K$ and $m_{_{d}}{=}0$ at $Q$. Vertical arrows indicate VB to CB transitions. Right: Joint optical density of states as a function of transition energy calculated for whole BZ.}
\end{figure}

We now consider tunneling from Mo atom to its 6 nearest neighbors $\vec{R}_{A1}-\vec{R}_{A6}$ Mo atoms as illustrated in Fig.~\ref{fig:F1}, with same for sulfur dimers. The second neighbor tunneling matrix elements are parameterized by tunneling amplitudes $W_{i}$ with dependence on $\vec{k}$ expressed by functions $g_i(\vec{k})$. Explicit forms of $g_{i}(\vec{k})$ and $W_i$ are given in the Appendix B. The explicit form of the Hamiltonian contains now dispersion of and coupling between $d-$ and $p-$ orbitals:
\begin{widetext}
\begin{eqnarray}
H\left(\vec{k} \right)=\left(\begin{array}{cccccc}
^{E_{m_{_{d}}=-2}}_{{+}W_{1}g_{0}(\vec{k})}& W_{3}g_{2}(\vec{k}) & W_{4}g_{4}(\vec{k}) & V_{1}f_{-1}(\vec{k}) & -V_{2}f_{0}(\vec{k}) & V_{3}f_{1}(\vec{k}) \\
& ^{E_{m_{_{d}}=0}}_{{+}W_{2}g_{0}(\vec{k})} & W_{3}g_{2}(\vec{k})   & -V_{4}f_{0}(\vec{k}) & -V_{5}f_{1}(\vec{k}) & -V_{4}f_{-1}(\vec{k}) \\
&  & ^{E_{m_{_{d}}=2}}_{{+}W_{1}g_{0}(\vec{k})}    & -V_{3}f_{1}(\vec{k}) & -V_{2}f_{-1}(\vec{k}) & V_{1}f_{0}(\vec{k}) \\
&  &     & ^{E_{m_{_{p}}=-1}}_{{+}W_{5}g_{0}(\vec{k})} & 0 & W_{7}g_{2}(\vec{k}) \\
&  &     &  & ^{E_{m_{_{p}}=0}}_{{+}W_{6}g_{0}(\vec{k})} & 0 \\
&  &     &  &  & ^{E_{m_{_{p}}=1}}_{{+}W_{5}g_{0}(\vec{k})} \\
\end{array} \right)
\label{eq9}
\end{eqnarray}
\end{widetext}
Fig.~\ref{fig:F4} shows the energy bands obtained using first and second neighbor Hamiltonian, Eq.~\ref{eq9}, black squares, and ab--initio energy bands without spin--orbit (SO) coupling. We see that the gap opens up across the entire BZ due to direct interaction of $d$--orbitals. The right hand side of the figure shows excellent agreement of ab--initio and TB, Eq.~\ref{eq9}, conduction (CB) and valence (VB) energy bands. In particular, we see the second minimum in the CB at $Q$ point. The origin of the minimum at $Q$ point is analyzed in the left panel of Fig.~\ref{fig:F5} where different colors mark contributions from different $d$-orbitals. The size of circles denotes the contribution of different orbitals. At the $K$ point the top of the VB is composed mainly of $m_{_d}{=}2$ orbital and bottom of CB has $m_{_{d}}{=}0$ character. At $\Gamma$ point top of the VB has $m_{_{d}}{=}0$ character and bottom of the CB has $m_{_{d}}{=}{\pm}2$ character.  Hence the higher energy band with $m_{_{d}}{=}{-}2$ character has to cross the $m_{_{d}}{=}0$ CB. The crossing of $m_{_{d}}{=}0$ and $m_{_{d}}{=}{-}2$ bands leads to a maximum in the conduction band followed by a second minimum at $Q$.  Around the minimum at $Q$ point the conduction and valence bands are parallel. The nesting of CB and VB leads to a maximum in joint optical density of states, shown in the right panel of Fig.~\ref{fig:F5} and discussed already by, e.g., Castro--Neto and co--workers \cite{R17}.
\begin{figure}
\includegraphics[width=0.40\textwidth]{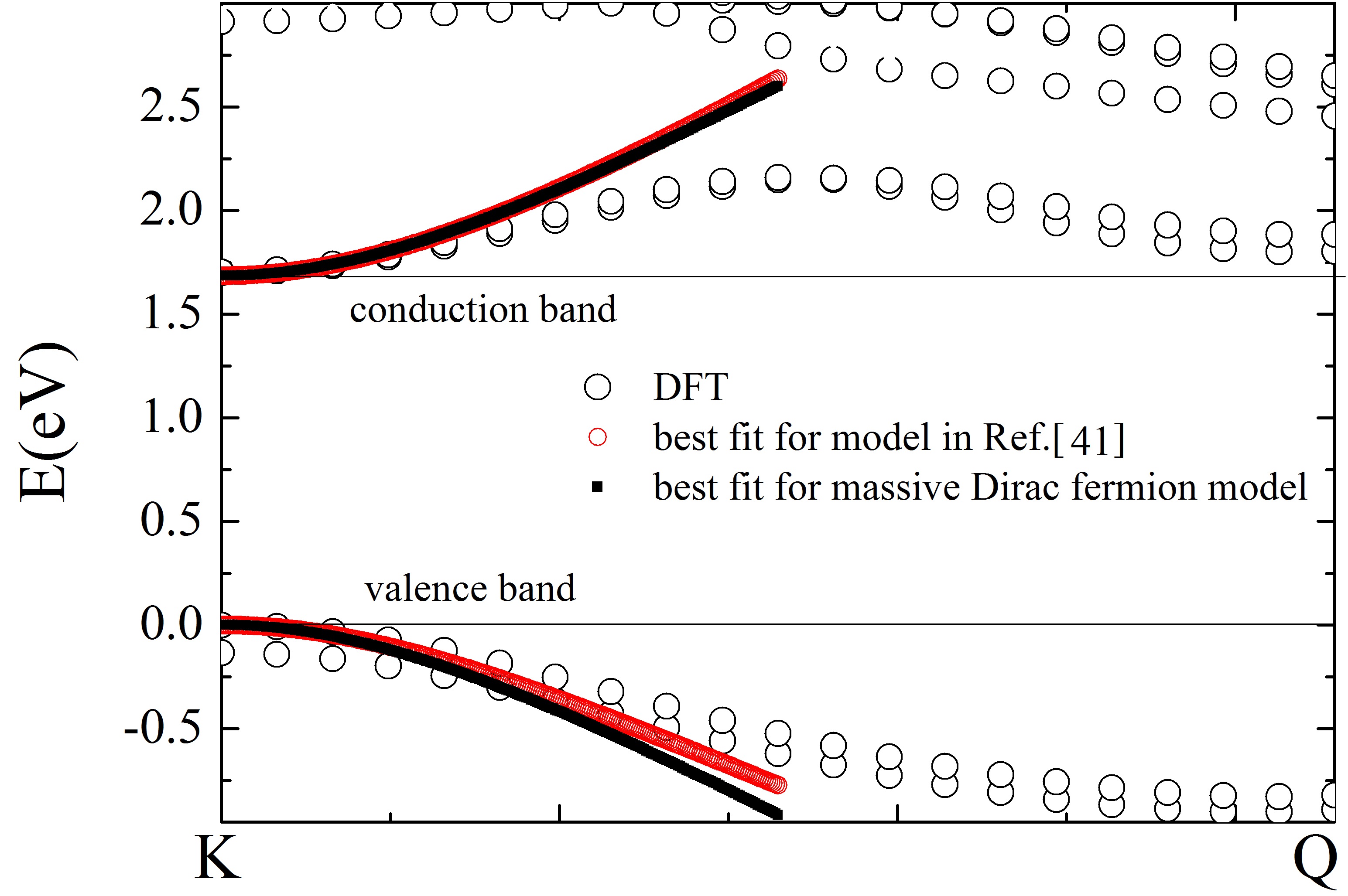}
\caption{\label{fig:F6}: Energy dispersion of effective massive Dirac Fermion model in the vicinity of $K$ point: DFT (empty circles), ref.[\onlinecite{R41}]-red circles, this work-black squares.}
\end{figure}

\section{Effective two-band massive Dirac fermion model}
With the 6-band model understood we now proceed to fit our results to the two--band massive Dirac Fermion model applicable in the vicinity of $K$ points. Following \textcite{R41} we write our two--band Hamiltonian $H_{2B}$ as a function of deviation $q$ from the wavector $k=K+q$ as:
\begin{displaymath}
H_{2B}(\vec{k})=a{\cdot} t\left(\begin{array}{cc}
 & \tau q_{x}{-}iq_{y} \\
\tau q_{x}{+}iq_{y} &  \\
\end{array} \right) {+}\frac{\Delta}{2}\left(\begin{array}{cc}
{1} &  \\
 & {-1} \\
\end{array} \right)
\end{displaymath}
\begin{equation}
{+}\left(\begin{array}{cc}
\alpha q^{2} &  \\
 & \beta q^{2} \\
\end{array} \right){+}\kappa\left(\begin{array}{cc}
& q_{+}^{2}  \\
q_{-}^{2} &  \\
\end{array} \right){-}\tau\frac{\eta}{2}q^{2}\left(\begin{array}{cc}
& q_{+}  \\
q_{-} &  \\
\end{array} \right)
\label{eq10}
\end{equation}
where $q_{\pm}{=}q_{x}\pm q_{y}$ and $\tau{=}{\pm}1$ for $K$,$-K$ valley's. Fig.~\ref{fig:F6} shows the results of fitting eigenenergies of Eq.~\ref{eq10} to our ab--initio results and results obtained by $k\cdot p$ theory of \textcite{R41}. We see a good agreement of all three results. The two--band model parameters used in Fig.~\ref{fig:F6} are $a{=}3.193$ \AA, $t{=}1.4111$ eV, $\Delta{=}1.6850$ eV, $\alpha{=}0.8341$ eV\AA$^{2}$, $\beta{=}0.8066$ eV\AA$^{2}$, $\kappa{=}{-}0.0354$ eV\AA$^{2}$, $\eta{=}{-}0.0833$ eV\AA$^{3}$. For $\alpha{=} \beta{=} \kappa{=} \eta{=}0$ Eq.~\ref{eq10} reduces to a massive Dirac Fermion model proposed by Xiao et al.[8] for the description of conduction and valence bands close to the $K$ point. Note that wavevector $k$ is measured from the $K$ point. Best parameters for massive Dirac Fermion model are $a{=}1.46$ \AA, $t{=}1.4677$ eV and $\Delta{=}1.6848$ eV.

\section{Magnetic field - Lande g-factors  }
We now describe response of TMDC's to the applied magnetic field \cite{R11, R23, R35, R38, R42, R43, R44, R45, R46, R47, R48, R49, R50, R51, R52, R53, R54, R57}. The perpendicular magnetic field $B$ couples to the orbital angular momentum $L$ as $H_{2}=\mu_{B}B_{z}L_{z}$. From symmetry analysis at the $K$ point the wavefunctions of conduction band are composed of $m_{_{d}}{=}0$ and $m_{_{p}}{=}{-}1$ orbitals as:
\begin{equation}
\begin{split}
&\Psi_{CB}^{\uparrow}\left(\vec{K},\vec{r}\right)=\\
&\frac{A_{m_{_{d}}=0}^{\vec{k}=\vec{K}}(CB)}{\sqrt{N_{UC}}}\sum_{i=1}^{N_{UC}}e^{i\vec{K}\cdot\vec{R}_{A,i}}\varphi_{l=2,m_{_{d}}=0}\left(\vec{r}-\vec{R}_{A,i}\right)+\\
&\frac{B_{m_{_{p}}=-1}^{\vec{k}=\vec{K}}(CB)}{\sqrt{N_{UC}}}\sum_{i=1}^{N_{UC}}e^{i\vec{K}\cdot\vec{R}_{B,i}}\varphi_{l=1,m_{_{p}}=-1}\left(\vec{r}-\vec{R}_{B,i}\right)
\end{split}
\label{eq11}
\end{equation}
With details of the analysis found in the Appendix C the energy of electron in CB at $K$ point is given by the contributions from the $m_{_{d}}{=}0$ orbital, equal to zero, and finite contribution from $m_{_{p}}{=}{-}1$ orbital as
\begin{displaymath}
\begin{split}
E_{CB}\left(\vec{K}\right)=&\bra{\Psi_{CB}^{\uparrow}\left(\vec{K}\right)}\hat{L}_{z}/\hbar\ket{\Psi_{CB}^{\uparrow}\left(\vec{K}\right)}\mu_{B}B_{z}=\\
&(-1)\left |B_{m_{_{p}}=-1}^{\vec{k}=\vec{K}}(CB)  \right|^{2}\mu_{B}B_{z}.
\end{split}
\end{displaymath}
At $-K$ point the energy of electron in CB is given by the contributions from the $m_{_{d}}{=}0$ orbital (no contribution) and contribution from $m_{_{p}}{=}{+}1$ orbital as
\begin{displaymath}
\begin{split}
E_{CB}\left(-\vec{K}\right)=&\bra{\Psi_{CB}^{\uparrow}\left(-\vec{K}\right)}\hat{L}_{z}/\hbar\ket{\Psi_{CB}^{\uparrow}\left(-\vec{K}\right)}\mu_{B}B_{z}=\\
&(+1)\left |B_{m_{_{p}}=+1}^{\vec{k}=-\vec{K}}(CB)  \right|^{2}\mu_{B}B_{z}.
\end{split}
\end{displaymath}
The valley Lande energy splitting $\Delta^{\textrm{CB}}_{\textrm{VL}}$ in the conduction band is given by
\begin{equation}
\begin{split}
&\Delta^{\textrm{CB}}_{\textrm{VL}}=E_{\textrm{CB}}({+}\vec{K}){-}E_{\textrm{CB}}({-}\vec{K})=\\
&\bigg[{-}1\left |B_{m_{_{p}}=-1}^{\vec{k}=\vec{K}}(\textrm{CB})  \right|^{2}{-}1\left|B_{m_{_{p}}=1}^{\vec{k}=-\vec{K}}(\textrm{CB}) \right|^{2}\bigg]\mu_{B}B_{z}=\\
&(-2)\left |B_{m_{_{p}}=-1}^{\vec{k}=\vec{K}}(\textrm{CB})  \right|^{2}\mu_{B}B_{z},
\end{split}
\label{eq12}
\end{equation}
where we used the fact that orbital compositions of $m_{p}{=}{\pm} 1$ orbitals at $K$ and ${-}K$ are equal. A similar analysis carried out for the valley Lande energy splitting $\Delta^{\textrm{VB}}_{\textrm{VL}}$ in the valence band gives $\Delta^{\textrm{VB}}_{\textrm{VL}}=2\bigg(2\left |A_{m_{_{d}}=2}^{\vec{k}=\vec{K}}(\textrm{VB})  \right|^{2}{+}1\left |B_{m_{_{p}}=+1}^{\vec{k}=\vec{K}}(\textrm{VB})  \right|^{2}\bigg)\mu_{B}B_{z}$. Using results from the 6 band model, Eq.~\ref{eq9}, gives the effective Lande g--factors of $g^{\textrm{CB}}_{\textrm{VL}}=-2|B_{m_{_{p}}=-1}^{\vec{k}=\vec{K}}(\textrm{CB})|^{2}=-0.4$ in the conduction band and $\Delta^{\textrm{VB}}_{\textrm{VL}}=2\bigg(2\left |A_{m_{_{d}}=2}^{\vec{k}=\vec{K}}(\textrm{VB})  \right|^{2}{+}1\left |B_{m_{_{p}}=+1}^{\vec{k}=\vec{K}}(\textrm{VB})  \right|^{2}\bigg)\mu_{B}B_{z}= 3.996$. By comparison, values deduced from Ref.[\onlinecite{R34}] give $g^{\textrm{CB}}_{\textrm{VL}}=-0.88, g^{\textrm{VB}}_{\textrm{VL}}=3.20$ and those from Ref.[\onlinecite{R32}] give $g^{\textrm{CB}}_{\textrm{VL}}=-0.24, g^{\textrm{VB}}_{\textrm{VL}}=3.44$.

\section{Magnetic field - valley Zeeman and Landau g-factors}
\begin{figure}
\includegraphics[width=0.50\textwidth]{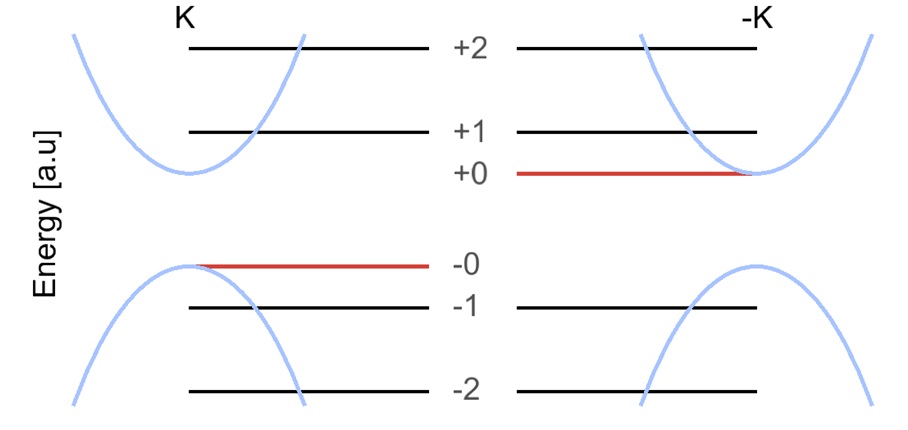}
\caption{\label{fig:F7}: Landau energy levels at $K$ and $-K$ points. Note the splitting of the (0)-energy level -- it is attached to the top of the valence band at $K$ and to the bottom of the CB at $-K$ point. The energy difference of Landau  $n{=}{+}1$ level at ${+}K$ and $n{=}0$ Landau level at $-K$ gives Valley Zeeman splitting.}
\end{figure}

We now discuss valley Zeeman splitting due to Landau quantization. We start with the massive Dirac Hamiltonian for $K$ point derived in Eq.~\ref{eq10}:
\begin{equation}
H_{2B}(\vec{k})=\frac{\Delta}{2}\left(\begin{array}{cc}
{1} &  \\
 & {-1} \\
\end{array} \right)+v_{F}\left(\begin{array}{cc}
 & \tau q_{x}{-}iq_{y} \\
\tau q_{x}{+}iq_{y} &  \\
\end{array} \right)
\label{eq13}
\end{equation}
With magnetic field $\vec{B}=B\hat{z}$ in the symmetric gauge, vector potential $\vec{A}{=}B/2(-y,x,0)$. We substitute $\vec{q}{\to} \vec{q}{+}e/c\vec{A}$, measure length in units of magnetic length $r{\to} r/l_{0}$, where $l_{0}{=}\sqrt{eB/c}$. Transformation into creation and annihilation operators \cite{R55}
\begin{equation}
\begin{split}
&\hat{a}^{\dagger}=\frac{1}{\sqrt{2}}\left(-\partial_{x}-i\partial_{y}+\frac{1}{2}(x+iy) \right),\\ &\hat{a}=\frac{1}{\sqrt{2}}\left(\partial_{x}-i\partial_{y}+\frac{1}{2}(x-iy) \right)
\end{split}
\label{eq14}
\end{equation}
gives massive Dirac Fermion Hamiltonian in magnetic field as
\begin{equation}
H_{\textrm{m.DF}}(\vec{k})=\frac{\Delta}{2}\left(\begin{array}{cc}
{1} &  \\
 & {-1} \\
\end{array} \right)+v\left(\begin{array}{cc}
 & -i\hat{a} \\
+i\hat{a}^{\dagger} &  \\
\end{array} \right)
\label{eq15}
\end{equation}
where $v{=}\sqrt{2}v_{F}/l_{0}$. The eigenfunctions $\Psi_{n,m}^{C/V}{=}\left(\alpha_{n}^{C/V}\ket{n{-}1,m},\beta_{n}^{C/V}\ket{n,m}\right)^{T}$ of Hamiltonian, Eq.~\ref{eq15}, are spinors in the basis of CB and VB states at $K$ ($-K$), with eigenenergies of the n$^{\textrm{th}}$ Landau level $E_{n}^{C/V}{=}\pm\sqrt{(\Delta/2)^{2}+v^{2}n}$. Here, massive Dirac Fermion nature manifests itself in eigenvectors expressed as a combination of states with different $n$, which differs for both valleys. The energy spectrum contains three types of states for $K$ and $-K$ points: positive (negative) energies with $n\geq1$ for conduction (valence) band states indicated by indices C(V) and $n=0$ Landau level (0LL) in each valley. The key result \cite{R42,R43} is that in the $K$ valley the 0LL is attached to the top of the valence band (negative energy) while in the $-K$ valley 0LL is attached to the bottom of the conduction band (positive energy). Fig.~\ref{fig:F7} shows the energy spectrum for $K$ and  $-K$ valleys, with $n=0$ LL shown in red. 

The valley Zeeman splitting in the conduction band is given by the energy difference between electron in ${+}K$ valley, $E_{1}(+K){=}\sqrt{\left(\Delta/2\right)^{2}+\left(\sqrt{2}v_{F}/l_{0}\right)^{2}}$ and electron in the $-K$ valley $E_{0}(-K){=}\sqrt{\left(\Delta/2\right)^{2}+\left(\sqrt{2}v_{F}/l_{0}\right)^{2}}$:
\begin{equation}
\begin{split}
\Delta_{VZ}=&E_{1}^{C}(K)-E_{0}^{C}(-K)=\frac{\Delta}{2}\left(\sqrt{1+\left(\frac{2v}{\Delta l_{0}}\right)^{2}}-1\right)\approx \\
&\frac{\Delta}{2}\left(\frac{\sqrt{2}v_{F}}{\Delta l_{0}}\right)^{2}=\Delta\left(\frac{v_{F}}{\Delta l_{0}}\right)^{2}\hbar \omega_{c},
\end{split}
\label{eq16}
 \end{equation}
where $\hbar \omega_{c}$ is the cyclotron energy. We see that valley Zeeman splitting is proportional to the cyclotron energy and the ratio of Fermi velocity to the energy gap \cite{R57}. 

We can now compare the valley Lande and Zeeman contributions for MoS$_{2}$. For magnetic field $B {=}1$ T, we obtain the following values of splitting:
\begin{equation}
\Delta_{\textrm{VL}}^{\textrm{CB}}=-2\left|B^{\textrm{CB}}_{m_{_{p}}=-1}(K)\right|^{2}\mu_{B}B=-0.023 \textrm{ meV} 
\label{eq17}
\end{equation}
\begin{equation}
\Delta_{\textrm{VL}}^{\textrm{VB}}=2\left(2\left|A^{\textrm{CB}}_{m_{_{d}}=2}\right|^{2}{+}1\left|B^{\textrm{CB}}_{m_{_{p}}=-1}\right|^{2}\right)\mu_{B}B=0.231 \textrm{ meV} 
\label{eq18}
\end{equation}
\begin{equation}
\Delta_{\textrm{VZ}}=\Delta\left(\frac{v_{F}}{\Delta}\right)^{2}\hbar \omega_{c}=\Delta\left(\frac{v_{F}}{\Delta}\right)^{2}\hbar \frac{eB}{m_{0}}=1.395 \textrm{ meV}
\label{eq19}
\end{equation}
First two values are in excellent agreement with recently reported \cite{R44} experimental Lande splitting of approximately $0.230$ meV $T^{-1}$.

We now discuss the effect of spin-orbit interaction on the Landau level spectrum. The Hamiltonian for both spin down and up at the $K$ point can be written as:
\begin{eqnarray}
H=\left(\begin{array}{cccc}
\frac{\Delta}{2}{-}\frac{\Delta^{\textrm{C}}_{\textrm{SO}}}{2} & -iv\hat{a} &  & \\
iv\hat{a}^{\dagger}&{-}\frac{\Delta}{2}{-}\frac{\Delta^{\textrm{V}}_{\textrm{SO}}}{2}  &  & \\
  & &\frac{\Delta}{2}{+}\frac{\Delta^{\textrm{C}}_{\textrm{SO}}}{2} & -iv\hat{a}\\
  & & iv\hat{a}^{\dagger}&{-}\frac{\Delta}{2}{+}\frac{\Delta^{\textrm{V}}_{\textrm{SO}}}{2} \\
\end{array} \right),
\label{eq20}
\end{eqnarray}
where $\Delta^{\textrm{C/V}}_{\textrm{SO}}$ is the spin splitting for conduction (valence) band. In analogy with Eq.~\ref{eq15} we obtain the eigenvectors $\Psi_{n,m}^{\pm,\sigma}{=}\left(\alpha_{n,\sigma}^{\textrm{C/V},\Delta_{\textrm{SO}}}\ket{n{-}1,m},\beta_{n,\sigma}^{\textrm{C/V},\Delta_{\textrm{SO}}}\ket{n,m}\right)^{T}$ for $K$ and eigenvectors $\Psi_{n,m}^{\pm,\sigma}{=}\left(\left(\alpha_{n,\sigma}^{\textrm{C/V},-\Delta_{\textrm{SO}}}\right)^{*}\ket{n,m},\beta_{n,\sigma}^{\textrm{C/V},-\Delta_{\textrm{SO}}}\ket{n{-}1,m}\right)^{T}$ for the ${-}K$ valley. The corresponding  eigenvalues are given by $E_{n,\sigma}^{\textrm{C/V}}=\sigma(\Delta_{\textrm{SO}}^{\textrm{C}}+\Delta_{\textrm{SO}}^{\textrm{V}})/4\pm\sqrt{\left(\Delta+\sigma\left(\Delta_{\textrm{SO}}^{\textrm{C}}-\Delta_{\textrm{SO}}^{\textrm{V}}\right)/2\right)^{2}/4+v^{2}n}$, where $\sigma{=}{\pm}1$ for spin up or down. The LL spectrum becomes even more asymmetric between the valleys. Because of the interaction between valence and conduction band the strong SO coupling in the valence band leads to spin splitting in the conduction band.

\section{Conclusions}
We presented here a tight--binding theory of transition metal dichalcogenides. We derived an effective tight binding Hamiltonian and elucidated the electron tunneling from metal to  dichalcogenides orbitals at different points of the BZ. This allowed us to discuss the band gaps at $K$ points in the BZ, the origin of secondary conduction band minima at $Q$ points and their role in band nesting and strong light matter interaction. The Lande and Zeeman valley splitting as well as the effective mass Dirac Fermion Hamiltonian in the magnetic field was determined. 

\begin{acknowledgments}
MB acknowledges financial support from National Science Center (NCN), Poland, grant Sonata No. 2013/11/D/ST3/02703. L.SZ and P.H. acknowledge support from NSERC and uOttawa University Research Chair in Quantum Theory of Materials, Nanostructures and Devices. One of us (PH) thanks J. Kono, S. Crooker, A. Stier and M. Potemski for discussions.
\end{acknowledgments}

\appendix
\section{Nearest neighbor matrix elements}
Matrix elements of nearest neighbor tunneling Hamiltonian, Eq.~\ref{eq8}, are expressed by $k$-independent parameters~$V_{i}$ 
\begin{displaymath}
V_{1}=\frac{1}{\sqrt{2}}\left[\frac{\sqrt{3}}{2}\left(\frac{d^{2}_{\bot}}{d^{2}}-1\right)V_{dp\sigma}-\left(\frac{d^{2}_{\bot}}{d^{2}}+1\right)V_{dp\pi} \right]\frac{d_{\|}}{d},
\end{displaymath}
\begin{displaymath}
V_{2}=\frac{1}{2}\left[\sqrt{3}V_{dp\sigma}-2V_{dp\pi} \right]\frac{d_{\bot}}{d}\left(\frac{d_{\|}}{d}\right)^{2},
\end{displaymath}
\begin{displaymath}
V_{3}=\frac{1}{\sqrt{2}}\left[\frac{\sqrt{3}}{2}V_{dp\sigma}-V_{dp\pi}\right]\left(\frac{d_{\|}}{d}\right)^{3},
\end{displaymath}
\begin{displaymath}
V_{4}=\frac{1}{2}\left[\left(3\frac{d^{2}_{\bot}}{d^{2}}-1\right)V_{dp\sigma}-2\sqrt{3}\frac{d^{2}_{\bot}}{d^{2}}V_{dp\pi} \right]\frac{d_{\|}}{d},
\end{displaymath}
\begin{equation}
V_{5}=\frac{1}{2}\left[\left(3\frac{d^{2}_{\bot}}{d^{2}}-1\right)V_{dp\sigma}-2\sqrt{3}\left(\frac{d^{2}_{\bot}}{d^{2}}-1\right)V_{dp\pi} \right]\frac{d_{\bot}}{d}
\end{equation}
and $k$-dependent factors $(k_{x}d_{\|}\to k_{x},k_{y}d_{\|}\to k_{y} )$
\begin{equation*}
\begin{split}
f_{0}(\vec{k})=e^{ik_{x}}&+e^{-ik_{x}/2}e^{i\sqrt{3}k_{y}/2}e^{-i2\pi/3}\\&+e^{-ik_{x}/2}e^{-i\sqrt{3}k_{y}/2}e^{i2\pi/3},\\
\end{split}
\end{equation*}
\begin{equation*}
\begin{split}
f_{-1}(\vec{k})=e^{ik_{x}}&+e^{-ik_{x}/2}e^{i\sqrt{3}k_{y}/2}e^{i2\pi/3}\\&+e^{-ik_{x}/2}e^{-i\sqrt{3}k_{y}/2}e^{-i2\pi/3},\\
\end{split}
\end{equation*}
\begin{equation}
f_{+1}(\vec{k})=e^{ik_{x}}+e^{-ik_{x}/2}e^{i\sqrt{3}k_{y}/2}+e^{-ik_{x}/2}e^{-i\sqrt{3}k_{y}/2}.
\end{equation}

\section{Next nearest neighbor matrix elements}
Parameters of the second neighbor tunneling in Eq.~\ref{eq9} are given by $k$-independent terms $W_{i}$
\begin{equation}
\begin{split}
&W_{1}=\frac{1}{8}\left(3V_{dd\sigma}+4V_{dd\pi}+V_{dd\delta}\right),\\
&W_{2}=\frac{1}{4}\left(V_{dd\sigma}+3V_{dd\delta}\right),\\
&W_{3}=-\frac{\sqrt{3}}{4\sqrt{2}}\left(V_{dd\sigma}-V_{dd\delta}\right),\\
&W_{4}=\frac{1}{8}\left(3V_{dd\sigma}-4V_{dd\pi}+V_{dd\delta}\right),\\
&W_{5}=\frac{1}{2}\left(V_{pp\sigma}+V_{pp\pi}\right),\\
&W_{6}=V_{pp\pi},\\ 
&W_{7}=\frac{1}{2}\left(V_{pp\sigma}-V_{pp\pi}\right),\\
\end{split}
\end{equation}
and $k$-dependent functions $g_{i}$
\begin{equation*}
g_{0}(\vec{k})=4\cos{\left(3k_{x}/2\right)}\cos{\left(\sqrt{3}k_{y}/2\right)}+2\cos{\left(\sqrt{3}k_{y}\right)},
\end{equation*}
\begin{equation*}
\begin{split}
g_{2}(\vec{k})=-2\cos{\left(\sqrt{3}k_{y}\right)}&+2\cos{\left(3k_{x}/2+\sqrt{3}k_{y}/2 \right)e^{i\pi/3}}\\
&+2\cos{\left(3k_{x}/2-\sqrt{3}k_{y}/2 \right)e^{-i\pi/3}}
\end{split}
\end{equation*}
\begin{equation}
\begin{split}
g_{4}(\vec{k})=2\cos{\left(\sqrt{3}k_{y}\right)}&+2\cos{\left(3k_{x}/2+\sqrt{3}k_{y}/2 \right)e^{i2\pi/3}}\\
&+2\cos{\left(3k_{x}/2-\sqrt{3}k_{y}/2 \right)e^{-i2\pi/3}}
\end{split}
\end{equation}
Slater-Koster parameters found by fitting our second nearest neighbor model to DFT bandstructure used to create Fig.~\ref{fig:F4} and Fig.~\ref{fig:F5} are given in Table~\ref{TAB1}.
\begin{table}[h]
	\centering
		\begin{tabular}{|c|c|c|c|}
		\hline
			{parameter} & {best fit (in eV)} & {parameter} & {best fit (in eV)} \\ \hline
			$E_{m_{_d}=0,\pm2}$ & -0.03 &$V_{dd\sigma}$ &-1.10 \\ \hline
			$E_{m_{_p}=\pm 1}$ & -3.36 &$V_{dd\pi}$ &0.76 \\ \hline
			$E_{m_{_p}=0}$ & -4.78 &$V_{dd\delta}$ &0.27 \\ \hline
			$V_{dp\sigma}$ & -3.39 &$V_{pp\sigma}$ &1.19 \\ \hline
			$V_{dp\pi}$ & 1.10 &$V_{pp\pi}$ &-0.83 \\ \hline
\end{tabular}
\caption{\label{TAB1}Slater-Koster parameters fitted to DFT bandstructure.}
\end{table}

\onecolumngrid
\section{Lande g--factor}
To calculate Lande g--factor in perpendicular magnetic field $B_{z}$ we first analyze expectation value of $L_{z}$ operator for wavefunctions of A and B sublattices written as:
\begin{equation}
\begin{split}
&\Psi_{A}^{m}\left(\vec{k},\vec{r}\right)=\frac{1}{\sqrt{N_{UC}}}\sum_{i=1}^{N_{UC}}e^{i\vec{k}\cdot\vec{R}_{A,i}}\varphi_{m}\left(\vec{r}-\vec{R}_{A,i}\right),\\
&\Psi_{B}^{l}\left(\vec{k},\vec{r}\right)=\frac{1}{\sqrt{N_{UC}}}\sum_{j=1}^{N_{UC}}e^{i\vec{k}\cdot\vec{R}_{B,j}}\varphi_{l}\left(\vec{r}-\vec{R}_{B,j}\right).
\end{split}
\end{equation}
For $L_{z}=-i\hbar\left(\vec{r}\times\vec{\nabla}_{\vec{r}}\right)_{z}$ we have therefore
\begin{equation}
\bra{\Psi_{B}^{l}\left(\vec{k},\vec{r}\right)}(-i\hbar)\left(\vec{r}\times\vec{\nabla}_{\vec{r}}\right)_{z}\ket{\Psi_{A}^{m}\left(\vec{k},\vec{r}\right)}=\frac{-i\hbar}{N_{UC}}\sum_{i,j=1}^{N_{UC}}\int d\vec{r}e^{i\vec{k}\cdot\vec{R}_{AB}} \varphi^{*}_{l}\left(\vec{r}-\vec{R}_{B,j}\right) \left(\vec{r}\times\vec{\nabla}_{\vec{r}}\right)_{z}\varphi_{m}\left( \vec{r}-\vec{R}_{A,i}\right),
\end{equation}
where $\vec{R}_{AB}=\vec{R}_{A,i}-\vec{R}_{B,j}$. To evaluate Eq. (C2) we introduce new variables $\vec{u}_{i}=\vec{r}-\vec{R}_{A,i}$ to analyze the action of operator $\vec{\nabla}_{\vec{r}}$ on orbitals localized at $\vec{R}_{A,i}$
\begin{equation}
\left(\vec{r}\times \frac{\partial}{\partial \vec{r}} \right)\varphi_{m}\left( \vec{r}-\vec{R}_{A,i}\right)=m\varphi_{m}\left(\vec{u}_{i}\right)+\left(\vec{R}_{A,i}\times \vec{p}_{u_{i}}\right)_{z}\varphi_{m}\left(\vec{u}_{i}\right).
\end{equation}
Using this and shifting variables on sites B as $\vec{r}-\vec{R}_{B,j}=\vec{u}_{i}+\vec{R}_{A,i}-\vec{R}_{B,j}$ we obtain
\begin{equation}
\begin{split}
\bra{\Psi_{B}^{l}\left(\vec{k},\vec{r}\right)}(-i\hbar)\left(\vec{r}\times\vec{\nabla}_{\vec{r}}\right)_{z}\ket{\Psi_{A}^{m}\left(\vec{k},\vec{r}\right)}=&
m\frac{(-i\hbar)}{N_{UC}}\sum_{\vec{R}_{AB},\vec{R}_{A}}^{N_{UC}}\int d\vec{u}e^{i\vec{k}\cdot\vec{R}_{AB}}\underbrace{\varphi^{*}_{l}\left(\vec{u}+\vec{R}_{AB}\right)\varphi_{m}\left(\vec{u}\right)}_{\neq 0 \textrm{ only for }\vec{R}_{AB}=0}+\\
&\frac{(-i\hbar)}{N_{UC}}\sum_{\vec{R}_{AB},\vec{R}_{A}}^{N_{UC}}\int d\vec{u}e^{i\vec{k}\cdot\vec{R}_{AB}}
\varphi^{*}_{l}\left(\vec{u}+\vec{R}_{AB}\right)\left(\vec{R}_{A}\times\vec{p}_{\vec{u}_{i}}\right)_{z}\varphi_{m}\left(\vec{u}\right)
\end{split}
\end{equation}
First term on RHS of Eq. (C4) can be more transparently written as
\begin{equation}
m\frac{(-i\hbar)}{N_{UC}}\sum_{\vec{R}_{A}}^{N_{UC}}\underbrace{\int d\vec{u}\varphi^{*}_{l}\left(\vec{u}\right)\varphi_{m}\left(\vec{u}\right)}_{\delta_{lm}}=m\frac{(-i\hbar)}{N_{UC}}\sum_{\vec{R}_{A}}^{N_{UC}}\delta_{lm}=m\frac{(-i\hbar)}{N_{UC}}N_{UC}\delta_{lm}=(-i\hbar)m\delta_{lm}
\end{equation}
while the second term vanishes
\begin{equation}
\frac{(-i\hbar)}{N_{UC}}\sum_{\vec{R}_{A}}^{N_{UC}}\bigg[\vec{R}_{A}\underbrace{\sum_{\vec{R}_{AB}}^{N_{UC}}e^{i\vec{k}\cdot\vec{R}_{AB}}\int d\vec{u}\varphi^{*}_{l}\left(\vec{u}+\vec{R}_{AB}\right)\left(\vec{p}_{\vec{u}_{i}}\right)_{z}\varphi_{m}\left(\vec{u}\right)}_{\vec{R}_{A}\textrm{ independent}}\bigg]=0,
\end{equation}
because the sum over $\vec{R}_{A}$ is taken over an isotropic system. Finally, we get
\begin{equation}
\bra{\Psi_{B}^{l}\left(\vec{k},\vec{r}\right)}(-i\hbar)\left(\vec{r}\times\vec{\nabla}_{\vec{r}}\right)_{z}\ket{\Psi_{A}^{m}\left(\vec{k},\vec{r}\right)}=(-i\hbar)m\delta_{lm},
\end{equation}
which is used to calculate Lande energy splitting, e.g. for conduction band $E_{CB}(\pm K)=\left(\mp 1\right)\left|B_{m_{_{p}}=\mp 1}^{\vec{k}=K}(CB)\right|^{2}\mu_{B}B_{z}$. Analogous analysis can be performed for the valence band.
\twocolumngrid
\bibliography{apssamp}

\end{document}